\documentclass{article}
\usepackage[final,nonatbib]{nips_2018}
\usepackage[utf8]{inputenc} % allow utf-8 input
\usepackage[T1]{fontenc}    % use 8-bit T1 fonts
\usepackage{hyperref}       % hyperlinks
\usepackage{url}            % simple URL typesetting
\usepackage{booktabs}       % professional-quality tables
\usepackage{amsfonts}       % blackboard math symbols
\usepackage{nicefrac}       % compact symbols for 1/2, etc.
\usepackage{microtype}      % microtypography
\usepackage{enumitem}
\usepackage{graphicx}
\usepackage{amsmath}

\title{Automatic Documentation of ICD Codes with Far-Field Speech Recognition}

\author{
 Albert Haque \\
  Stanford University\\
  Stanford, CA 94305 \\
    \And
 Corinna Fukushima \\
  Rush University Medical Center\\
  Chicago, IL 60612 \\
}

\begin{document}
    
\maketitle

\begin{abstract}
Documentation errors increase healthcare costs and cause unnecessary patient deaths. As the standard language for diagnoses and billing, ICD codes serve as the foundation for medical documentation worldwide. Despite the prevalence of electronic medical records, hospitals still witness high levels of ICD miscoding. In this paper, we propose to automatically document ICD codes with far-field speech recognition. Far-field speech occurs when the microphone is located several meters from the source, as is common with smart homes and security systems. Our method combines acoustic signal processing with recurrent neural networks to recognize and document ICD codes in real time. To evaluate our model, we collected a far-field speech dataset of ICD-10 codes and found our model to achieve 87\% accuracy with a BLEU score of 85\%. By sampling from an unsupervised medical language model, our method is able to outperform existing methods. Overall, this work shows the potential of automatic speech recognition to provide efficient, accurate, and cost-effective healthcare documentation.
\end{abstract}

\section{Introduction}

More than 250,000 people die every year in the United States due to medical errors, making it the third leading cause of death \cite{james2013new, cdc2016deaths}.
Unsurprisingly, many of these errors are preventable, such as inaccurate drug doses, unlisted allergies, and wrong-site amputations.
Directly responsible for many of these errors is poor documentation \cite{hartel2011high, mulloy2008wrong, henderson2006quality}.
Worldwide, ICD codes serve as the standard language for diagnoses, treatment, and billing  \cite{icd2018usa, icd2017uk}.
However, ICD miscoding occurs as much as as 20\%, with similar rates dating back to the 1990s \cite{henderson2006quality, macintyre1997accuracy}.
Despite electronic medical records, documentation errors still occur and cost the United States up to \$25 billion each year \cite{lang2007consultant, farkas2008automatic}.

One of the main sources of ICD miscoding is during patient admission \cite{o2005measuring}.
This is paramount for emergency departments, which handle 141 million visits each year in the United States \cite{cdc2014edstats}.
However, emergency departments are overcrowded, leading to overworked and rushed clinical teams, which can ultimately produce more errors \cite{us2011gao, sun2013effect, hooper2010compassion}.
While there has been work on inferring ICD codes from text, those methods still require a written document \cite{larkey1995automatic, subotin2016method, tutubalina2017encoder, huang2018empirical}.
If we can automate such documentation, not only can we potentially reduce medical errors, but we can also free up time from medical teams, who spend up to 26\% of their time on documentation tasks alone \cite{ammenwerth2009time, arndt2017tethered}.

\begin{figure}[t]
    \centering
    \includegraphics[width=0.8\linewidth]{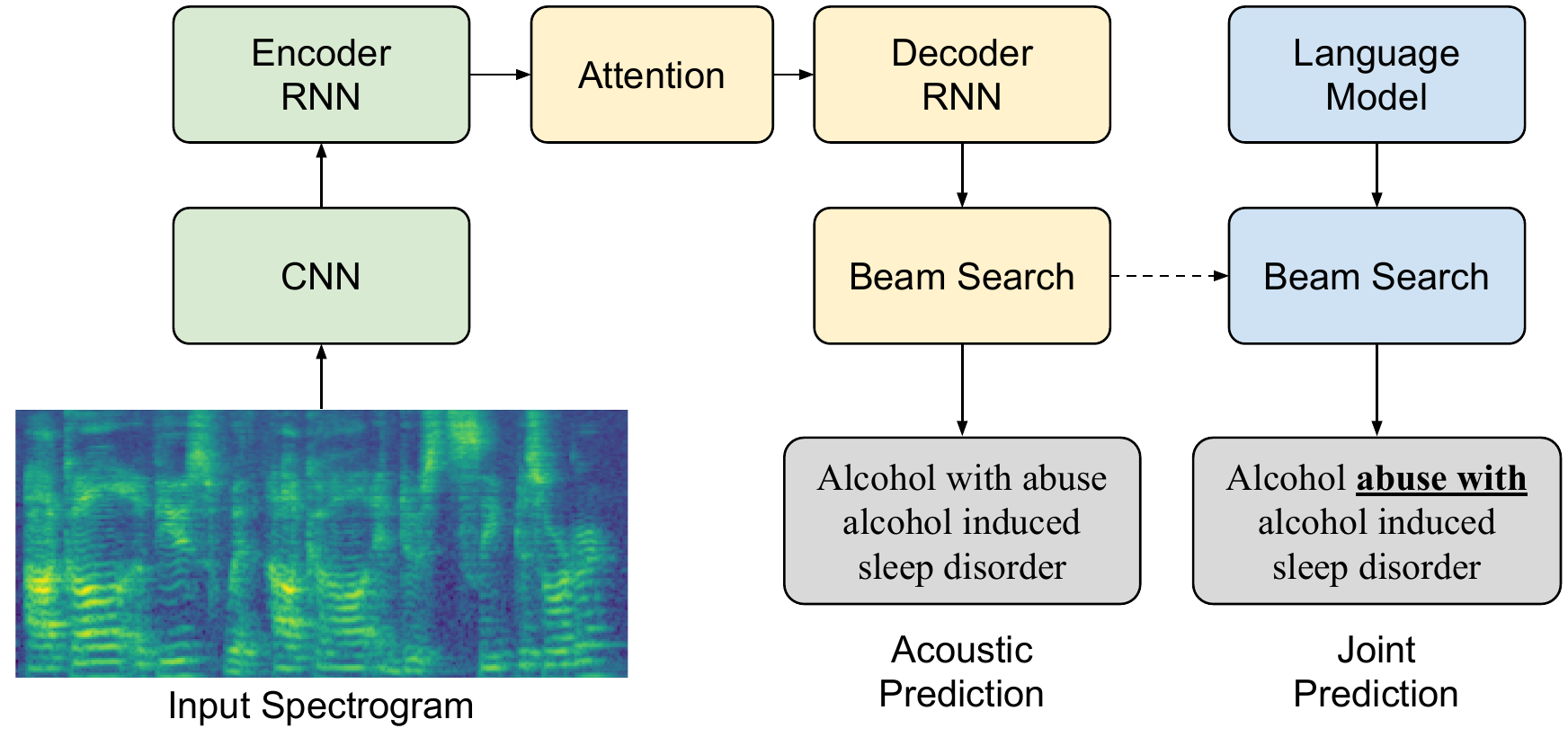}
    \caption{\textbf{Overview of our method.} The dashed arrow denotes random sampling. Green denotes the encoder, yellow is the decoder, blue is the language model, and gray is the model output.}
    \label{fig:method}
\end{figure}

Outside of healthcare, smart homes have enjoyed the benefits of voice-based home assistants such as Google Home and Amazon Alexa  \cite{li2017acoustic}.
These devices can perform interactive search queries to accelerate ordinary household workflows such as cooking or morning routines.
In this paper, we bring the advances of smart homes and far-field speech recognition to automatically document ICD codes in healthcare.
The input to our model is a spectrogram and the output is a text transcription \cite{chiu2018speech}.
By sampling from an unsupervised language model during training, we can achieve a knowledge distillation effect to improve overall transcription performance.

\section{Proposed Method}\label{sec:method}

Overall, our method consists of a sequence-to-sequence deep neural network as an acoustic model (see Figure \ref{fig:method}).
During training, the acoustic model samples from an unsupervised medical language model to improve transcription.

\textbf{Acoustic Model.}
Medical words are often longer than conversational speech.
They can contain many syllables, such as the words \textit{hyperventilation}.
As a consequence, the resulting ICD code is long, sometimes five to ten seconds in length.
Modeling the full spectrogram would require unrolling of the encoder RNN for an infeasibily large number of timesteps, on the order of hundreds to thousands of RNN time steps \cite{sainath2015learning}.
Even with truncated backpropagation, this would be a challenging task \cite{haykin2001kalman}.

Inspired by \cite{sainath2015learning}, we propose to reduce the temporal length of the input spectrogram by using a learned convolutional filter bank.
As shown in Figure \ref{fig:method}, silence appears throughout the spectrogram, denoted by dark blue regions.
By treating the spectrogram as an image, convolutional filters can not only reduce the temporal dimension but also compress redundant frequency-level information, such as the absence of high-frequencies.
However, we found convolutional networks insufficient to reduce the temporal dimension to a manageable length.
WaveNet, a method for speech synthesis, employs dilated convolutions to control the temporal receptive field at each layer of the network \cite{van2016wavenet, dutilleux1990implementation, yu2015multi}.
As a result, a single temporal representation from a high-level layer can encode hundreds, if not thousands of RNN timesteps.
Formally, let $h_i^j$ denote the hidden state of a long short-term memory (LSTM) \cite{hochreiter1997long} cell at the $i$-th timestep of the $j$-th layer (Equation \ref{eq:pblstm}).
For a pyramidal LSTM (pLSTM), outputs from the preceding layer, containing high-resolution temporal information, are concatenated,
\begin{equation}\label{eq:pblstm}
h_i^j = \textrm{LSTM}(h_{i-1}^j, h_i^{j-1}) \textrm{\quad and \quad}
    h_i^j = \textrm{pLSTM}\left(h_{i-1}^j, \left[ h_{\beta i}^{j-1}, h_{\beta i+1}^{j-1}, \cdots, h_{\beta (i+1) - 1}^{j-1} \right]\right)
\end{equation}
where $\left[ \cdot \right]$ denotes the concatenation operator.
In Equation \ref{eq:pblstm}, the output of a pLSTM unit is now a function of not only its previous hidden state, but also the outputs from $\beta$ previous timesteps from the layer below.
The pLSTM provides us with an exponential reduction in number of RNN timesteps.
Not only does the pyramidal RNN provide higher-level temporal features, but it also reduces the inference complexity \cite{chan2016listen}.
Because we do not use a bidirectional RNN, our model can run in real-time.
After the pLSTM (encoder) processes the input, the states are given to the decoder.

The concept of \textit{attention} has proven useful in many tasks \cite{chorowski2015attention, mascharka2018transparency}.
With attention, each step of the model's decoder has access to all of the encoder's outputs.
The goal of such an ``information shortcut" is to address the challenge of learning long-range temporal dependencies \cite{bengio1994learning}.
The distribution for the predicted word $y_i$ is a function of the decoder state $s_i$ and attention context $c_i$.
The context vector $c_i$ is produced by an attention mechanism \cite{chan2016listen}. Specifically, $c_i = \sum_u \alpha_{i,u}$,
where attention is defined as the alignment between the current decoder timestep $i$ and encoder timestep $j$:
\begin{equation}
    \alpha_{i,j} = \frac{\exp(e_{i,j})}{\sum_{j} \exp(e_{i,j})} \textrm{\quad where \quad} e_{i,j} = w^\top \tanh(Ws_{i-1} + Vh_j + b)
\end{equation}
and where the score between the output of the encoder or the hidden states, $h_j$, and the previous state of the decoder cell, $s_{i-1}$ is computed with $e_{i,u} = \langle \phi(s_i), \varphi(h_u)\rangle$ where $\phi$ and $\varphi$ are sub-networks, e.g. multi-layer perceptrons.
The $w$, $W$ and $V$ are learnable parameters.

The final output from the decoder is a sequence of word embeddings, which encode the transcribed sentence (e.g., one-hot vector).
This can be trained by optimizing the cross-entropy loss objective.
Many previous works end at this point by performing greedy-search or beam-search decoding methods \cite{sutskever2014sequence, graves2006connectionist}, but we propose to randomly sample from the language model to improve the transcription.

\textbf{Language Model.}
Our language model is an n-gram model.
Given a sequence of words $w_1, \cdots, w_n$, the model assigns a probability $p(w_1, w_2, \cdots , w_n)$ of the sequence occurring in the training corpus:
\begin{align}
    p(w_1, \cdots , w_n) &= \prod\limits_{i=1}^{n} p(w_i | w_1, \cdots, w_{i-1}) \\ 
    &\approx  \prod\limits_{i=1}^{n} \Big[ \lambda_1 p(w_i | w_{1}, \cdots, w_{i-1})+ \cdots + \lambda_{i-1} p(w_i | w_{i-1}) + \lambda_i p(w_i) \Big]  \label{eq:ngram}
\end{align}
where $\sum_i \lambda_i = 1$.
By default, $\lambda_i = 1/n$.
To overcome the paucity of higher-order n-grams, we approximate the n-gram probability by interpolating the individual n-gram probabilities.

\textbf{Combining the Acoustic and Language Models.}
To improve our ICD transcriptions, we impose a training constraint on the acoustic model, subject to the language model.
However, such a concept is not new.
When a language model is used during inference but not training \cite{chorowski2016towards, wu2016google}, we call this \textit{shallow fusion} .
This was extended by \cite{sriram2017cold} in the paper \textit{Cold Fusion}, to include the language model during training \cite{kannan2017analysis}.
However, in their work, the language model was fixed during training and kept as a feature extractor.
In our work, we keep the language model constant, but instead combine cold fusion with scheduled sampling \cite{bengio2015scheduled}.
The effect is a type of unsupervised knowledge distillation \cite{hinton2015distilling}.
We can combine the language model and our acoustic model to select the optimal transcription $\hat{y}^*$:
\begin{equation}\label{eq:loss}
    \hat{y}^* =  \operatorname*{argmin}_{\hat{y}} \left[ \frac{\lambda_\alpha}{\lambda_\alpha+\lambda_\ell} \log p_{\alpha}(\hat{y}|x) + \frac{\lambda_\ell}{\lambda_\alpha+\lambda_\ell} \log p_{\ell}(\hat{y}) \right]
\end{equation}
\noindent where $\alpha$ and $\ell$ denote the acoustic and language model, respectively.
The $\lambda$'s control the language model sampling probability and also serve as mixing hyperparameters.
The $p(\cdot)$'s denotes the posterior probability and $x$ denotes the input spectrogram.
The entire process is differentiable and can be optimized with first-order methods \cite{kingma2014adam}.

\section{Experiments}

\textbf{Dataset.}
We collected a far-field speech dataset of common ICD-10 codes used in emergency departments \cite{aapc2014icd}.
First, a pre-determined list of one hundred ICD-10 codes and descriptions was selected.
This resulted in 141 unique words with an average character length of 7.3, median length of 7, maximum length of 16, and a standard deviation of 3.2 characters.
Second, the vocabulary list was repeated five times by multiple speakers.
This was done to collect diverse pitch tracks and intonation from each speaker.
Third, full ICD code descriptions were generated by procedurally concatenating each individual word.
For a single ICD code, there were 20,730 acoustic variations per speaker.

A total of six speakers participated in data collection.
Each speaker stood 12 feet (3.6 meters) from a computer monitor and microphone.
For all experiments, one speaker was excluded from the training set and used as the test set.
This was done six times, such that each speaker was part of the test set once, and the average WER and BLEU are reported.
Although our procedural dataset generation can allow for millions of training examples, we limited the variations per ICD code to 1,000.
As a result, the training set consisted of 60,480 sentences per speaker, for a total training size of 302,400 sentences.
Words were converted to a one-hot word embedding and used as decoder targets.

\textbf{Language Model.}
The language model was trained on the entire ICD medical corpus \cite{icd2018usa}.
The corpus consisted of 94,127 sentences, 7,153 unique words, 922,201 total words.
Punctuations such as commas, dashes, and semi-colons were removed.
The model was trained on n-grams up to length 10.

\textbf{Metrics.}
We use word error rate (WER) and BLEU as metrics.
The WER is defined as $(S+D+I)/N$ where $S$ denotes the number of word substitutions, $D$ denotes deletions, $I$ denotes insertions, and $N$ denotes the number of words in the ground truth sentence. 
If $I=0$ then accuracy is equivalent to recall (i.e., sensitivity).
Word-level accuracy is denoted by $1-\textrm{WER}$.
Common in machine translation but less used in speech recognition, we use the Bilingual Evaluation Understudy (BLEU) metric since it can better capture contextual and syntactic roles of a word \cite{he2011word}.

\begin{table}[t]
    \centering
    \small
    \vspace{-4mm}
    \begin{tabular}{l|cc|cc}
    \toprule
         & \multicolumn{2}{c|}{WER $\downarrow$} & \multicolumn{2}{c}{BLEU $\uparrow$} \\
        Method & Native & Non-Native & Native & Non-Native \\ \midrule
        Human (Medically Trained)  & $2.4 \pm 1.2$ & $3.6 \pm 2.0$ & $95.9 \pm 0.7$ & $ 93.5 \pm 0.8$ \\
        Human (Untrained)  & $6.6 \pm 1.0$ & $8.0 \pm 1.4$ & $91.9 \pm 2.4$ & $ 90.6 \pm 2.5$  \\ \midrule
        Connectionist Temporal Classification \cite{graves2006connectionist} & $67.7 \pm 0.1$ & $70.2 \pm 0.2$  & $32.7 \pm 0.2$ & $ 28.4 \pm 0.2$  \\
        Sequence-to-Sequence \cite{sutskever2014sequence} & $32.1 \pm 0.1$ & $31.9 \pm 0.1$ & $66.8 \pm 0.1$   & $66.9 \pm 0.1$ \\ 
        Listen, Attend, and Spell \cite{chan2016listen, chiu2018speech} & $14.8 \pm 0.1$ & $21.4 \pm 0.2$ & $84.6 \pm 0.1$  & $80.5 \pm 0.2$ \\ 
        Cold Fusion \cite{deepspeech3} & $16.2 \pm 0.1 $ & $23.0 \pm 0.1$ & $82.9 \pm 0.1$ & $79.4 \pm 0.1$  \\ \midrule
        Our Method  & \textbf{$13.4 \pm 0.1$}  & \textbf{$ 19.1 \pm 0.1 $}  & $85.0 \pm 0.1$ &  $83.8 \pm 0.1$ \\
        \bottomrule
    \end{tabular}
    \caption{\textbf{Comparison with existing methods.} Lower WER and higher BLEU is better. Human refers to manual transcription. Native refers to native English speakers. All methods were trained and evaluated on our ICD-10 dataset. $\pm$ denotes the 95\% confidence interval.}
    \label{tab:sota}
\end{table}

\textbf{Results.}
Table \ref{tab:sota} shows quantitative results for existing methods and our proposed method.
For most methods, the performance on non-native English speakers is lower than native speakers.
This is to be expected due to the larger variances in non-native speech, especially for complex Latin medical words.
Surprisingly, Cold Fusion has a higher WER and lower BLEU than the LAS model, despite cold fusion using an external language model.
One explanation is the establishment of a dependence, on a potentially biased language model.
Our method could be viewed as the same as cold fusion, but with a smaller mixing parameter.
However, we did not run exhaustive tuning to find the optimal hyperparemeters for this dataset.
As for CTC, the poor performance could partially be attributed to CTC's design for phoneme-level recognition.
In our task, we evaluate CTC at word-level, which significantly increases the branching factor (i.e., there are more unique words than phonemes).

\textbf{Qualitative Comparison.}
Table \ref{tab:qualitative} shows ``qualitative" results of our model compared to the baselines.
While only two ICD codes are shown in Table \ref{tab:qualitative}, in general, many mistakes made by Cold Fusion and our method occur on words which have a complementary or opposite pair (e.g., with/without, left/right, lower/upper).
Either word from the pair is valid, according to the language model.
The slightest acoustic variation such as a breath or pause may cause the word to be incorrectly substituted.

\begin{table}[t]
    \centering
    \small
    \begin{tabular}{l|l|l} \toprule
    Method & Transcription & Transcription\\ \midrule
    CTC & \textbf{Generalized pain} & \textbf{Pain abdominal}  \\
    Seq2Seq & \textbf{Abdominal} pain &  \textbf{Pain} injury without loss of consciousness \\
    LAS & \textbf{Lower} abdominal pain & Intracranial injury without loss of consciousness \\
    Cold Fusion & Generalized abdominal pain & Intracranial injury \textbf{with} loss of consciousness  \\
    Our Method & Generalized abdominal pain & Intracranial injury without loss of consciousness \\ \midrule
    Ground Truth & Generalized abdominal pain & Intracranial injury without loss of consciousness \\ \bottomrule
\end{tabular}
\caption{\textbf{Transcriptions for two ICD codes.} Bold text indicates a substitution, insertion, or deletion error. Each column indicates a single test-set example. The ground truth is shown in the bottom row.}
\label{tab:qualitative}
\end{table}

\section{Conclusion}

In this work, we presented a method to automatically document ICD codes with far-field speech recognition.
Our method combines acoustic signal processing techniques with deep learning-based approaches to recognize ICD codes.
There is future research to be done on handling multiple speakers and tonal languages such as Chinese.
Recent work on style and speaker tokens may prove beneficial \cite{wang2018style, haque2018conditional}.
Overall, this work shows the potential of modern automatic speech recognition to provide efficient, accurate, and cost-effective healthcare documentation.

\clearpage
\newpage
\small
\bibliographystyle{abbrv}
\bibliography{main}

\begin{thebibliography}{10}

\bibitem{aapc2014icd}
{AAPC}.
\newblock Icd-10 top 50 codes in emergency departments, 2014.

\bibitem{ammenwerth2009time}
E.~Ammenwerth and H.-P. Sp{\"o}tl.
\newblock The time needed for clinical documentation versus direct patient
  care.
\newblock {\em Methods of Information in Medicine}, 2009.

\bibitem{arndt2017tethered}
B.~G. Arndt, J.~W. Beasley, M.~D. Watkinson, J.~L. Temte, W.-J. Tuan, C.~A.
  Sinsky, and V.~J. Gilchrist.
\newblock Tethered to the ehr: primary care physician workload assessment using
  ehr event log data and time-motion observations.
\newblock {\em The Annals of Family Medicine}, 2017.

\bibitem{deepspeech3}
E.~Battenberg, J.~Chen, R.~Child, A.~Coates, Y.~Gaur, Y.~Li, H.~Liu,
  S.~Satheesh, D.~Seetapun, A.~Sriram, and Z.~Zhu.
\newblock Exploring neural transducers for end-to-end speech recognition.
\newblock {\em Automatic Speech Recognition and Understanding Workshop}, 2017.

\bibitem{bengio2015scheduled}
S.~Bengio, O.~Vinyals, N.~Jaitly, and N.~Shazeer.
\newblock Scheduled sampling for sequence prediction with recurrent neural
  networks.
\newblock In {\em Neural Information Processing Systems}, 2015.

\bibitem{bengio1994learning}
Y.~Bengio, P.~Simard, and P.~Frasconi.
\newblock Learning long-term dependencies with gradient descent is difficult.
\newblock {\em Trans. on Neural Networks}, 1994.

\bibitem{cdc2014edstats}
{CDC}.
\newblock National hospital ambulatory medical care survey: 2014 emergency
  department summary tables, 2014.

\bibitem{cdc2016deaths}
{CDC}.
\newblock Deaths and mortality, 2016.

\bibitem{chan2016listen}
W.~Chan, N.~Jaitly, Q.~Le, and O.~Vinyals.
\newblock Listen, attend and spell: A neural network for large vocabulary
  conversational speech recognition.
\newblock In {\em International Conference on Acoustics, Speech, and Signal
  Processing}, 2016.

\bibitem{chorowski2016towards}
J.~Chorowski and N.~Jaitly.
\newblock Towards better decoding and language model integration in sequence to
  sequence models.
\newblock {\em arXiv}, 2016.

\bibitem{chorowski2015attention}
J.~K. Chorowski, D.~Bahdanau, D.~Serdyuk, K.~Cho, and Y.~Bengio.
\newblock Attention-based models for speech recognition.
\newblock In {\em Neural Information Processing Systems}, 2015.

\bibitem{icd2018usa}
{CMS}.
\newblock Icd-10-cm official guidelines for coding and reporting, 2016.

\bibitem{dutilleux1990implementation}
P.~Dutilleux.
\newblock An implementation of the “algorithme {\`a} trous” to compute the
  wavelet transform.
\newblock In {\em Wavelets}. Springer, 1990.

\bibitem{farkas2008automatic}
R.~Farkas and G.~Szarvas.
\newblock Automatic construction of rule-based icd-9-cm coding systems.
\newblock In {\em BMC Bioinformatics}, 2008.

\bibitem{graves2006connectionist}
A.~Graves, S.~Fern{\'a}ndez, F.~Gomez, and J.~Schmidhuber.
\newblock Connectionist temporal classification: labelling unsegmented sequence
  data with recurrent neural networks.
\newblock In {\em International Conference on Machine Learning}, 2006.

\bibitem{haque2018conditional}
A.~Haque, M.~Guo, and P.~Verma.
\newblock Conditional end-to-end audio transforms.
\newblock {\em Interspeech}, 2018.

\bibitem{hartel2011high}
M.~J. Hartel, L.~P. Staub, C.~R{\"o}der, and S.~Eggli.
\newblock High incidence of medication documentation errors in a swiss
  university hospital due to the handwritten prescription process.
\newblock {\em BMC Health Services Research}, 2011.

\bibitem{haykin2001kalman}
S.~S. Haykin et~al.
\newblock {\em Kalman Filtering and Neural Networks}.
\newblock Wiley Online Library, 2001.

\bibitem{he2011word}
X.~He, L.~Deng, and A.~Acero.
\newblock Why word error rate is not a good metric for speech recognizer
  training for the speech translation task?
\newblock In {\em International Conference on Acoustics, Speech, and Signal
  Processing}, 2011.

\bibitem{henderson2006quality}
T.~Henderson, J.~Shepheard, and V.~Sundararajan.
\newblock Quality of diagnosis and procedure coding in icd-10 administrative
  data.
\newblock {\em Medical Care}, 2006.

\bibitem{hinton2015distilling}
G.~Hinton, O.~Vinyals, and J.~Dean.
\newblock Distilling the knowledge in a neural network.
\newblock {\em arXiv}, 2015.

\bibitem{hochreiter1997long}
S.~Hochreiter and J.~Schmidhuber.
\newblock Long short-term memory.
\newblock {\em Neural Computation}, 1997.

\bibitem{hooper2010compassion}
C.~Hooper, J.~Craig, D.~R. Janvrin, M.~A. Wetsel, and E.~Reimels.
\newblock Compassion satisfaction, burnout, and compassion fatigue among
  emergency nurses compared with nurses in other selected inpatient
  specialties.
\newblock {\em Journal of Emergency Nursing}, 2010.

\bibitem{huang2018empirical}
J.~Huang, C.~Osorio, and L.~W. Sy.
\newblock An empirical evaluation of deep learning for icd-9 code assignment
  using mimic-iii clinical notes.
\newblock {\em arXiv}, 2018.

\bibitem{james2013new}
J.~T. James.
\newblock A new, evidence-based estimate of patient harms associated with
  hospital care.
\newblock {\em Journal of Patient Safety}, 2013.

\bibitem{chiu2018speech}
D.~Jaunzeikare, A.~Kannan, P.~Nguyen, H.~Sak, A.~Sankar, J.~Tansuwan, N.~Wan,
  Y.~Wu, and X.~Zhang.
\newblock Speech recognition for medical conversations.
\newblock {\em Interspeech}, 2018.

\bibitem{kannan2017analysis}
A.~Kannan, Y.~Wu, P.~Nguyen, T.~N. Sainath, Z.~Chen, and R.~Prabhavalkar.
\newblock An analysis of incorporating an external language model into a
  sequence-to-sequence model.
\newblock {\em arXiv}, 2017.

\bibitem{kingma2014adam}
D.~P. Kingma and J.~Ba.
\newblock Adam: A method for stochastic optimization.
\newblock {\em International Conference on Learning Representations}, 2015.

\bibitem{lang2007consultant}
D.~Lang.
\newblock Natural language processing in the health care industry.
\newblock {\em Cincinnati Children’s Hospital Medical Center, Winter}, 2007.

\bibitem{larkey1995automatic}
L.~S. Larkey and W.~B. Croft.
\newblock Automatic assignment of icd9 codes to discharge summaries.
\newblock Technical report, University of Massachusetts at Amherst, 1995.

\bibitem{li2017acoustic}
B.~Li, T.~Sainath, A.~Narayanan, J.~Caroselli, M.~Bacchiani, A.~Misra,
  I.~Shafran, H.~Sak, G.~Pundak, K.~Chin, et~al.
\newblock Acoustic modeling for google home.
\newblock {\em Interspeech}, 2017.

\bibitem{macintyre1997accuracy}
C.~R. MacIntyre, M.~J. Ackland, E.~J. Chandraraj, and J.~E. Pilla.
\newblock Accuracy of icd--9--cm codes in hospital morbidity data, victoria:
  implications for public health research.
\newblock {\em Australian and New Zealand Journal of Public Health}, 1997.

\bibitem{mascharka2018transparency}
D.~{Mascharka}, P.~{Tran}, R.~{Soklaski}, and A.~{Majumdar}.
\newblock Transparency by design: Closing the gap between performance and
  interpretability in visual reasoning.
\newblock {\em arXiv}, 2018.

\bibitem{mulloy2008wrong}
D.~F. Mulloy and R.~G. Hughes.
\newblock Wrong-site surgery: a preventable medical error.
\newblock {\em Patient Safety and Quality; An Evidence-Based Handbook for
  Nurses}, 2008.

\bibitem{icd2017uk}
{NHS}.
\newblock National clinical coding standards icd-10, 2017.

\bibitem{o2005measuring}
K.~J. O'malley, K.~F. Cook, M.~D. Price, K.~R. Wildes, J.~F. Hurdle, and C.~M.
  Ashton.
\newblock Measuring diagnoses: Icd code accuracy.
\newblock {\em Health Services Research}, 2005.

\bibitem{van2016wavenet}
A.~v.~d. Oord, S.~Dieleman, H.~Zen, K.~Simonyan, O.~Vinyals, A.~Graves,
  N.~Kalchbrenner, A.~Senior, and K.~Kavukcuoglu.
\newblock Wavenet: A generative model for raw audio.
\newblock {\em arXiv}, 2016.

\bibitem{sainath2015learning}
T.~N. Sainath, R.~J. Weiss, A.~Senior, K.~W. Wilson, and O.~Vinyals.
\newblock Learning the speech front-end with raw waveform cldnns.
\newblock In {\em Interspeech}, 2015.

\bibitem{sriram2017cold}
A.~Sriram, H.~Jun, S.~Satheesh, and A.~Coates.
\newblock Cold fusion: Training seq2seq models together with language models.
\newblock {\em arXiv}, 2017.

\bibitem{subotin2016method}
M.~Subotin and A.~R. Davis.
\newblock A method for modeling co-occurrence propensity of clinical codes with
  application to icd-10-pcs auto-coding.
\newblock {\em Journal of the American Medical Informatics Association}, 2016.

\bibitem{sun2013effect}
B.~C. Sun, R.~Y. Hsia, R.~E. Weiss, D.~Zingmond, L.-J. Liang, W.~Han,
  H.~McCreath, and S.~M. Asch.
\newblock Effect of emergency department crowding on outcomes of admitted
  patients.
\newblock {\em Annals of Emergency Medicine}, 61(6):605--611, 2013.

\bibitem{sutskever2014sequence}
I.~Sutskever, O.~Vinyals, and Q.~V. Le.
\newblock Sequence to sequence learning with neural networks.
\newblock In {\em Neural Information Processing Systems}, 2014.

\bibitem{tutubalina2017encoder}
E.~Tutubalina and Z.~Miftahutdinov.
\newblock An encoder-decoder model for icd-10 coding of death certificates.
\newblock {\em arXiv}, 2017.

\bibitem{us2011gao}
{US GAO}.
\newblock Hospital emergency departments: Crowding continues to occur, and some
  patients wait longer than recommended time frames.
\newblock {\em {Government Accountability Office}}, GAO-09-347, 2009.

\bibitem{wang2018style}
Y.~Wang, D.~Stanton, Y.~Zhang, R.~Skerry-Ryan, E.~Battenberg, J.~Shor, Y.~Xiao,
  F.~Ren, Y.~Jia, and R.~A. Saurous.
\newblock Style tokens: Unsupervised style modeling, control and transfer in
  end-to-end speech synthesis.
\newblock {\em arXiv}, 2018.

\bibitem{wu2016google}
Y.~Wu, M.~Schuster, Z.~Chen, Q.~V. Le, M.~Norouzi, W.~Macherey, M.~Krikun,
  Y.~Cao, Q.~Gao, K.~Macherey, et~al.
\newblock Google's neural machine translation system: Bridging the gap between
  human and machine translation.
\newblock {\em arXiv}, 2016.

\bibitem{yu2015multi}
F.~Yu and V.~Koltun.
\newblock Multi-scale context aggregation by dilated convolutions.
\newblock {\em arXiv}, 2015.

\end{thebibliography}

\end{document}